\documentclass[aps,prd,twocolumn,nopacs,floatfix,amsmath,nofootinbib,superscriptaddress,amssymb,floatfix]{revtex4}
\usepackage{graphicx,color,dcolumn,booktabs,bm}
\usepackage{longtable,lscape}
\usepackage{pdfpages}
\usepackage{txfonts}
\usepackage{overpic}
\usepackage{amssymb}
\usepackage{makecell}
\usepackage{indentfirst}
\usepackage{feynmf}
\usepackage{slashed}
\usepackage{cases}
\usepackage{color}
\usepackage{multirow}
\usepackage{longtable,lscape}
\usepackage{threeparttable}
\usepackage{epstopdf}
\usepackage{makecell}

\usepackage{ulem}
\usepackage{enumerate}
\makeatletter

\newcommand{\Rmnum}[1]{\expandafter\@slowromancap\romannumeral #1@}
\makeatother
\usepackage{threeparttable}
\usepackage{graphicx,color,dcolumn,booktabs,bm}
\usepackage[colorlinks,
            citecolor=blue,
            anchorcolor=red,
            menucolor=red,
            linkcolor=red,
            filecolor=red,
            runcolor=red,
            urlcolor=blue,
            frenchlinks=true]{hyperref}
\usepackage{epstopdf}
\usepackage{array}

\begin{document}
\title{Investigation of low-lying $\Omega_b(1P)$ states in an unquenched coupled-channel framework}

\author{Zi-Le Zhang}
\affiliation{College of Physics and Electronic Engineering, Nanyang Normal University, Nanyang 473061, China}

\author{Si-Qiang Luo}\email{luosq15@lzu.edu.cn}
\affiliation{School of Physical Science and Technology, Lanzhou University, Lanzhou 730000, China}
\affiliation{Lanzhou Center for Theoretical Physics, Key Laboratory of Theoretical Physics of Gansu Province, Key Laboratory of Quantum Theory and Applications of MoE, Gansu Provincial Research Center for Basic Disciplines of Quantum Physics, Lanzhou University, Lanzhou 730000, China}
\affiliation{MoE Frontiers Science Center for Rare Isotopes, Lanzhou University, Lanzhou 730000, China}
\affiliation{Research Center for Hadron and CSR Physics, Lanzhou University and Institute of Modern Physics of CAS, Lanzhou 730000, China}

\author{Shuai-Wei Wang}
\affiliation{College of Physics and Electronic Engineering, Nanyang Normal University, Nanyang 473061, China}

\author{Qin Chang}
\affiliation{Institute of Particle and Nuclear Physics, Henan Normal University, Xinxiang, Henan 453007, China}

\begin{abstract}
In this work, we study the unquenched effects on the low-lying $\Omega_b(1P)$ states with a coupled-channel equations. We reveal how the unquenched effects affect the $\Omega_b(1P)$ spectroscopy and component mixing. The numerical results indicate that the $J^P=1/2^-$ $\Omega_b(1P)$ state dominated by the $j_\ell=0$ configuration exhibits significant coupled-channel effects due to its $S$-wave coupling to the $\Xi_b\bar{K}$ channel, where $j_\ell$ denotes the total angular momentum of the light flavor degrees-of-freedom. In this scenario, the mass of this state may  be shifted close to or below the $\Xi_b\bar{K}$ threshold, making the radiative and isospin-breaking channels kinematically allowed decay processes. The present analysis provides a coupled-channel perspective on the low-lying $\Omega_b(1P)$ spectrum and offers guidance for future experimental studies of excited bottom baryons.
\end{abstract}

\maketitle

\section{Introduction}\label{sec1}

The spectroscopy of heavy baryons provides an important window into the hadron structure and strong interactions~\cite{Chen:2016spr,Guo:2017jvc,Cheng:2015iom,Brambilla:2019esw,Liu:2019zoy,Chen:2022asf,Dong:2021juy,Cheng:2021qpd,Bai:2026atm,Wang:2025dur,Liu:2024uxn}. In particular, singly heavy baryons are especially useful because they allow one to disentangle the role of heavy-quark symmetry from the internal dynamics of the light degrees of freedom \cite{Meng:2022ozq}. Among them, the $\Omega_b(bss)$ baryon is of special interest. Owing to its relatively simple flavor structure, the low-lying $\Omega_b$ excitations provide a clean system for investigating orbital excitations and testing quark model descriptions.

Experimentally, significant progress has been made in the $\Omega_b$ sector in recent years. In 2020, LHCb reported four narrow excited structures in the $\Xi_b^0K^-$ invariant-mass spectrum, namely $\Omega_b(6316)^-$, $\Omega_b(6330)^-$, $\Omega_b(6340)^-$, and $\Omega_b(6350)^-$~\cite{LHCb:2020tqd}. Their measured masses and widths are summarized in Table \ref{table:exp}. These states are commonly interpreted as candidates for the low-lying $\Omega_b(1P)$ excitations and have stimulated considerable theoretical interest. In conventional quark model analyses, they are usually assigned to the $\lambda$-mode $\Omega_b(1P)$ multiplet \cite{Wang:2017kfr,Xiao:2020oif,Liang:2020hbo,Yang:2020zrh,Karliner:2020fqe,Chen:2020mpy,Wang:2020pri,XuYongJiang:2020cht,Mao:2015gya,Mutuk:2020rzm,Luo:2024jov}. However, alternative studies have emphasized that meson-baryon dynamics molecular states may also play an important role in this energy region~\cite{Liang:2017ejq,Liang:2020dxr}. Since the currently available experimental information is still largely limited to masses and narrow widths, the internal structure of these states remains far from settled.

In the conventional quark model, the $\lambda$-mode excited $\Omega_{b}(1P)$ baryon are expected to comprise five states. Heavy-quark spin symmetry and strong-decay analyses suggest that four of them are narrow, while the $J^P=1/2^{-}$ state with $j_\ell=0$ should be relatively broad~\cite{Wang:2017kfr,Xiao:2020oif,Liang:2020hbo,Yang:2020zrh,Chen:2020mpy}. The four narrow states can naturally be associated with the structures observed by LHCb~\cite{LHCb:2020tqd}, in agreement with theoretical expectations~\cite{Xiao:2020oif}. In contrast, the remaining one could couple with $\Xi_b\bar{K}$ via $S$-wave and is therefore expected to have a large decay width. However, corresponding broad state still missed in the LHCb's experiment~\cite{LHCb:2020tqd}.

\begin{table}[htbp]\label{table:exp}
\caption{The experimental information on the masses and widths of the low-lying excited $\Omega_b$ states is taken from Ref. \cite{LHCb:2020tqd,ParticleDataGroup:2026mpi}.}
\label{tab:parameter}
\renewcommand\arraystretch{1.25}
\begin{tabular*}{86mm}{@{\extracolsep{\fill}}ccccc}
\toprule[1.00pt]
\toprule[1.00pt]
States             &Mass (MeV)                       &Width (MeV) \\
\midrule[0.75pt]
$\Omega_b(6316)^-$ &$6315.64\pm 0.31\pm 0.07\pm 0.50$&$<4.2$ \\
$\Omega_b(6330)^-$ &$6330.30\pm 0.28\pm 0.07\pm 0.50$&$<4.7$ \\
$\Omega_b(6340)^-$ &$6339.71\pm 0.26\pm 0.05\pm 0.50$&$<1.8$ \\
$\Omega_b(6350)^-$ &$6349.88\pm 0.35\pm 0.05 \pm 0.50$&$<3.2$ \\
\bottomrule[1.00pt]
\bottomrule[1.00pt]
\end{tabular*}
\end{table}

Such a situation is not unique to the $\Omega_b$ sector. Owing to heavy-quark flavor symmetry, the charmed counterpart is expected to share similar properties with $\Omega_b$. Both the LHCb~\cite{LHCb:2017uwr,LHCb:2021ptx,LHCb:2023sxp} and Belle~\cite{Belle:2017ext} Collaborations have observed several narrow structures in the $\Omega_c(1P)$ region, which are generally consistent with the predicted $\lambda$-mode excitations~\cite{Chen:2017sci,Karliner:2017kfm,Wang:2017hej,Wang:2017vnc,Padmanath:2017lng,Cheng:2017ove,Wang:2017zjw,Zhao:2017fov,Chen:2017gnu,Aliev:2017led,Agaev:2017lip}. Nevertheless, the $J^P=1/2^{-}$ state dominated by the $j_\ell=0$ configuration, which is expected to be broad due to its strong $S$-wave coupling to the $\Xi_c\bar{K}$ channel, remains unobserved in current experiments~\cite{LHCb:2017uwr,Belle:2017ext}.

To explore the possible origin of this discrepancy, Refs.~\cite{Luo:2021dvj,Zhang:2025gar} investigated the $\Omega_c$ spectrum within an unquenched quark model. Their results suggested that the strong interaction between the bare three-quark configuration and the nearby $\Xi_c\bar{K}$ channel may induce significant coupled-channel effects. As a consequence, the bare state can experience a substantial downward mass shift and may even be driven below the $\Xi_c\bar{K}$ threshold. Such a scenario is not unprecedented in hadron spectroscopy. The $\Lambda(1405)$, $D_{s0}^*(2317)$, $D_{s1}^{\prime}(2460)$, $X(3872)$, and $\Lambda_c(2940)$ have all been discussed as states potentially influenced by strong coupled-channel effects~\cite{Heikkila:1983wd,Ono:1983rd,Ono:1985jt,Ono:1985eu,Tornqvist:1984fy,Silvestre-Brac:1991qqx,Pennington:2007xr,Barnes:2007xu,Zhou:2011sp,Danilkin:2010cc,Liu:2016wxq,Zhang:2009bv,Ortega:2009hj,Li:2009ad}.

Motivated by the studies of $\Omega_c(1P)$, it is natural to ask whether a similar phenomenon also occurs in $\Omega_b(1P)$ owing to heavy quark flavor symmetry. We investigate how the coupling between bare quark-core configurations and nearby hadronic channels modifies the masses and internal composition of the physical $\Omega_b(1P)$ states. In particular, we focus on the coupled-channel effects arising from the coupling of the bare state associated with the missing state to the $\Xi_b\bar{K}$ channel. Our goal is to clarify the role of unquenched effect in bottom-baryon spectroscopy and to provide useful theoretical input for future experimental studies of the $\Omega_b$ system.

This article is organized as follows. Following the Introduction, we proceed to present the unquenched formalism for computing the masses of the low-lying $\Omega_b$ states in Sec.~\ref{sec2}. Subsequently, in Sec.~\ref{sec3}, we carry out the calculation of the relevant matrix elements. In Sec.~\ref{sec4}, we discuss the unquenched effects in the $\Omega_b(1P)$ sector. Finally, we conclude with a brief summary in Sec.~\ref{sec5}.

\section{The unquenched model} \label{sec2}

\begin{figure}
    \centering
    \includegraphics[width=0.25\textwidth]{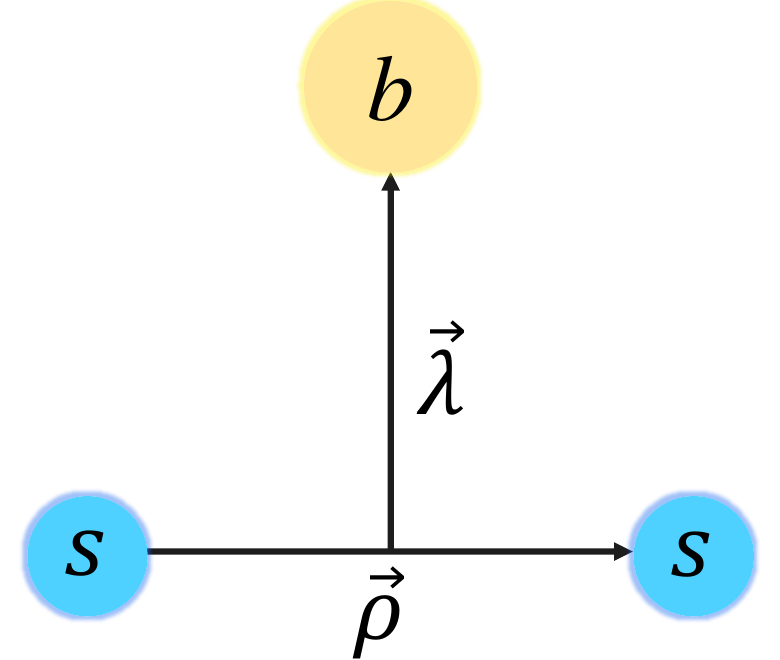}
    \caption{ The definitions of internal Jacobi coordinates $\vec{\lambda}$ and $\vec{\rho}$.}
    \label{fig:jacodi}
\end{figure}

As shown in Fig.~\ref{fig:jacodi}, within the framework of heavy-quark spin symmetry, and neglecting other degrees of freedom, the spin-spatial basis of a singly heavy flavor baryon could be written as
\begin{equation}\label{eq:JMJ}
|JM_J\rangle=|[[[s_{q_1}s_{q_2}]_{s_{12}}[n_\rho l_\rho n_\lambda l_\lambda]_L]_{j_\ell}s_{Q_3}]_{JM_J}\rangle,
\end{equation}
where $s_{q_1}$ and $s_{q_2}$ denote the spins of the two light quarks, while $s_{Q_3}$ denotes the spin of the heavy quark. As illustrated in Fig.~\ref{fig:jacodi}, the relative coordinate between the two light flavor quarks is the $\rho$-mode,  whereas the coordinate between the heavy flavor quarks and the center of mass of the two light flavor quarks is the $\lambda$-mode. The quantum numbers $n_\rho$ and $l_\rho$ denote the radial and orbital angular momentum quantum numbers associated with the $\rho$ mode, respectively, while $n_\lambda$ and $l_\lambda$ denote the corresponding quantum numbers for the $\lambda$ mode. $L$ is the total orbital angular momentum. $j_\ell$ is angular momentum of the light flavor degree-of-freedom.

Since the bare $\Omega_b(1P,\,1/2^-)$ states are the mixtures of the quark model basis states with $j_\ell=0$ and $1$, and the bare $\Omega_b(1P,\,3/2^-)$ states are the mixtures of the basis states with $j_\ell=1$ and $2$, respectively. They can be written as
\begin{equation}
\begin{pmatrix}
\left|\frac{1}{2}^{-}\right\rangle_1 \\
\left|\frac{1}{2}^{-}\right\rangle_2
\end{pmatrix}
=
\begin{pmatrix}
\cos\theta_\frac{1}{2} & -\sin\theta_\frac{1}{2} \\
\sin\theta_\frac{1}{2} &\cos\theta_\frac{1}{2}
\end{pmatrix}
\begin{pmatrix}
\left|j_\ell=0,\frac{1}{2}^{-}\right\rangle \\
\left|j_\ell=1,\frac{1}{2}^{-}\right\rangle
\end{pmatrix},
\end{equation}
and
\begin{equation}
\begin{pmatrix}
\left|\frac{3}{2}^{-}\right\rangle_1 \\
\left|\frac{3}{2}^{-}\right\rangle_2
\end{pmatrix}
=
\begin{pmatrix}
\cos\theta_\frac{3}{2} & -\sin\theta_\frac{3}{2} \\
\sin\theta_\frac{3}{2} & \cos\theta_\frac{3}{2}
\end{pmatrix}
\begin{pmatrix}
\left|j_\ell=1,\frac{3}{2}^{-}\right\rangle \\
\left|j_\ell=2,\frac{3}{2}^{-}\right\rangle
\end{pmatrix}.
\end{equation}
Here, $j_\ell$ is the quantum number of the light degrees of freedom in the $j\!-\!j$ coupling scheme. We use $\theta_J$ to denote the mixing angle of the physical state with total spin-parity $J^P$. In particular, $\theta_{1/2}$ and $\theta_{3/2}$ correspond to the mixing angle with $J^P=1/2^-$ and $J^P=3/2^-$, respectively. Accordingly, the unquenched dynamics should be described within a coupled-channel framework involving two bare states. In the following, we present the formalism explicitly.

A physical state involving two bare configurations and two hadron-hadron channels can be expanded as~\cite{Guo:2017jvc,Kalashnikova:2005ui,Danilkin:2009hr,Lu:2017hma,Anwar:2018yqm,Ortega:2009hj,Ortega:2016pgg,Ortega:2021fem,Ortega:2021yis,Lu:2016mbb}
\begin{eqnarray}
\begin{split}\label{eq:phy2}
\left|\Psi\right\rangle
=&\,c_{\alpha_1}\left|\Psi_{\alpha_1}\right\rangle
+c_{\alpha_2}\left|\Psi_{\alpha_2}\right\rangle
+\int {\rm d}^3\mathbf{p}\,\phi_{\beta_1}(\mathbf{p})\left|\beta_1,\mathbf{p}\right\rangle \\
&+\int {\rm d}^3\mathbf{p}\,\phi_{\beta_2}(\mathbf{p})\left|\beta_2,\mathbf{p}\right\rangle ,
\end{split}
\end{eqnarray}
with the normalization condition
\begin{equation}\label{eq:Norm}
|c_{\alpha_1}|^2+|c_{\alpha_2}|^2
+\int \left|\phi_{\beta_1}(\mathbf{p})\right|^2 {\rm d}^3\mathbf{p}
+\int \left|\phi_{\beta_2}(\mathbf{p})\right|^2 {\rm d}^3\mathbf{p}
=1,
\end{equation}
where $\alpha_1$ and $\alpha_2$ label the two bare $\Omega_b$ configurations, while $\beta_1$ and $\beta_2$ label the relevant hadron-hadron channels. The corresponding coupled-channel Schr\"odinger equation can then be written as \cite{Lu:2016mbb,Luo:2021dvj}
\begin{equation}\label{eq:coupchaSch}
\left(\begin{array}{cccc}
\hat{H}_0 & \hat{H}_0 & \hat{H}_{I} & \hat{H}_{I} \\
\hat{H}_0 & \hat{H}_0 & \hat{H}_{I} & \hat{H}_{I} \\
\hat{H}_{I} & \hat{H}_{I} & \hat{H}_{\beta_1} & 0 \\
\hat{H}_{I} & \hat{H}_{I} & 0 & \hat{H}_{\beta_2}
\end{array}\right)
\left(\begin{array}{c}
c_{\alpha_1}|\Psi_{\alpha_1}\rangle\\
c_{\alpha_2}|\Psi_{\alpha_2}\rangle\\
\phi_{\beta_1}(\mathbf{p}) \left|\beta_1,\mathbf{p} \right\rangle\\
\phi_{\beta_2}(\mathbf{p}) \left|\beta_2,\mathbf{p} \right\rangle
\end{array}\right)
=
M
\left(\begin{array}{c}
c_{\alpha_1}|\Psi_{\alpha_1}\rangle\\
c_{\alpha_2}|\Psi_{\alpha_2}\rangle\\
\phi_{\beta_1}(\mathbf{p}) \left|\beta_1,\mathbf{p} \right\rangle\\
\phi_{\beta_2}(\mathbf{p}) \left|\beta_2,\mathbf{p} \right\rangle
\end{array}\right),
\end{equation}
where $\hat H_0$ denotes the Hamiltonian of the bare $\Omega_b$ basis states, which can be obtained within the conventional potential model. $\hat H_I$ describes the coupling between the bare states and the hadron-hadron channels, while $\hat H_{\beta_1}$ and $\hat H_{\beta_2}$ represent the Hamiltonians of the two hadron-hadron channels, respectively.

To make the coupled-channel structure explicit, we project Eq.~(\ref{eq:coupchaSch}) onto the basis states $\langle \Psi_{\alpha_1}|$, $\langle \Psi_{\alpha_2}|$, $\langle \beta_1,\mathbf p|$, and $\langle \beta_2,\mathbf p|$. In this way, the full Schr\"odinger equation is reduced to a set of coupled equations for the bare state coefficients and the hadron-hadron channel wave functions,
\begin{equation}\label{eq:EXtwobareEQ1}
\begin{split}
	&\int \phi_{\beta_1}({\bf p})H_{\alpha_1\to \beta_1}^*({\bf p}){\rm d}^3{\bf p}+\int \phi_{\beta_2}({\bf p})H_{\alpha_1\to \beta_2}^*({\bf p}){\rm d}^3{\bf p}\\
	&+M_{\alpha_1}c_{\alpha_1}+M^{\prime}c_{\alpha_2}=M c_{\alpha_1},\\
	&\int \phi_{\beta_1}({\bf p})H_{\alpha_2\to \beta_1}^*({\bf p}){\rm d}^3{\bf p}+\int \phi_{\beta_2}({\bf p})H_{\alpha_2\to \beta_2}^*({\bf p}){\rm d}^3{\bf p}\\
	&+M^{\prime}c_{\alpha_1}+M_{\alpha_2}c_{\alpha_2}=M c_{\alpha_2},
 \end{split}
\end{equation}
and
\begin{equation}\label{eq:EXtwobareEQ2}
\begin{split}
	&c_{\alpha_1}H_{\alpha_1\to \beta_1}({\bf p})+c_{\alpha_2}H_{\alpha_2\to \beta_1}({\bf p})
	+E_{\beta_1}\phi_{\beta_1}({\bf p})=M \phi_{\beta_1}({\bf p}),\\
    &c_{\alpha_1}H_{\alpha_1\to \beta_2}({\bf p})+c_{\alpha_2}H_{\alpha_2\to \beta_2}({\bf p})
	+E_{\beta_2}\phi_{\beta_2}({\bf p})=M \phi_{\beta_2}({\bf p}),
\end{split}
\end{equation}
where
\begin{equation}
\begin{split}\label{eq:TranM}
 &M_{\alpha_1}=\langle\Psi_{\alpha_1}|\hat{H}_0|\Psi_{\alpha_1}\rangle,\\
 &M_{\alpha_2}=\langle\Psi_{\alpha_2}|\hat{H}_0|\Psi_{\alpha_2}\rangle,\\
 &M^{\prime}=\langle\Psi_{\alpha_1}|\hat{H}_0|\Psi_{\alpha_2}\rangle
 =\langle\Psi_{\alpha_2}|\hat{H}_0|\Psi_{\alpha_1}\rangle,\\
 \end{split}
\end{equation}
specifies the bare state masses and the off-diagonal mixing terms between different bare states. $E_{\beta}$ denotes the free part of the energy for the corresponding hadron-hadron channel $\beta$. The transition amplitude between the bare state $\alpha_i$ and the hadron-hadron channel $\beta_j$ is defined as
 \begin{equation}\label{eq:TrMChannel}
    \begin{split}
 &H_{\alpha_i\to \beta_j}({\bf p})=\langle \beta_j,{\bf p}|\hat{H}_{I}|{\Psi_{\alpha_i}}\rangle,
\end{split}
\end{equation}
where the indices $i$ and $j$ independently take the values $1$ and $2$. The same convention is adopted below.

Based on Eq. (\ref{eq:EXtwobareEQ2}), the probability amplitude for the hadron-hadron channel can be recast into the following expression,
\begin{equation}
    \begin{split}\label{eq:hadronCW}
        &\phi_{\beta_j}({\bf p})=\frac{c_{\alpha_1}H_{\alpha_1\to \beta_j}({\bf p})+c_{\alpha_2}H_{\alpha_2\to \beta_j}({\bf p})}{M-E_{\beta_j}},
    \end{split}
\end{equation}

Substituting Eqs. (\ref{eq:TrMChannel}) and (\ref{eq:hadronCW}) back into Eq.~(\ref{eq:EXtwobareEQ1}), the effects of the intermediate hadron-hadron channels can be encoded into a set of energy-dependent and mixing terms,
\begin{equation}
    \begin{split}
        &\Delta M^{\alpha_i\beta_j}=\int \frac{|H_{\alpha_i\to \beta_j}({\bf p})|^2}{M-E_{\beta_j}} {\rm d}^3{\bf p},\\
        &\Delta M^{\alpha_1\alpha_2\beta_j}=\int \frac{H_{\alpha_1\to \beta_j}^*({\bf p})H_{\alpha_2\to \beta_j}({\bf p})}{M-E_{\beta_j}} {\rm d}^3{\bf p}.
        \end{split}
\end{equation}

After inserting Eqs. (\ref{eq:TranM})-(\ref{eq:hadronCW}) into Eq.~(\ref{eq:EXtwobareEQ1}) and making use of the corresponding energy dependences, the coupled channel equations for the bare state coefficients can thus be rewritten as
\begin{widetext}
\begin{equation}
    \begin{split}
        &(M_{\alpha_1}+\Delta M^{\alpha_1\beta_1}+\Delta M^{\alpha_1\beta_2})c_{\alpha_1}
        +(M^{\prime}+\Delta M^{\alpha_1\alpha_2\beta_1}+\Delta M^{\alpha_1\alpha_2\beta_2})c_{\alpha_2}=Mc_{\alpha_1},\\
        &(M^{\prime}+\Delta M^{\alpha_1\alpha_2\beta_1}+\Delta M^{\alpha_1\alpha_2\beta_2})c_{\alpha_1}
        +(M_{\alpha_2}+\Delta M^{\alpha_2\beta_1}+\Delta M^{\alpha_2\beta_2})c_{\alpha_2}=Mc_{\alpha_2}.
    \end{split}
\end{equation}
Equivalently, one arrives at the following effective $2\times 2$ eigenvalue equation in the bare state space,
\begin{equation}\label{eq:multiequSch}
\left(\begin{array}{cc}
	M_{\alpha_1}+\Delta M^{\alpha_1\beta_1}+\Delta M^{\alpha_1\beta_2}&M^{\prime}+\Delta M^{\alpha_1\alpha_2\beta_1}+\Delta M^{\alpha_1\alpha_2\beta_2}\\
	M^{\prime}+\Delta M^{\alpha_1\alpha_2\beta_1}+\Delta M^{\alpha_1\alpha_2\beta_2}&M_{\alpha_2}+\Delta M^{\alpha_2\beta_1}+\Delta M^{\alpha_2\beta_2}
\end{array}\right)
\left(\begin{array}{c}
c_{\alpha_1}\\
c_{\alpha_2}
\end{array}\right)=M\left(\begin{array}{c}
c_{\alpha_1}\\
c_{\alpha_2}
\end{array}\right).
\end{equation}
\end{widetext}
By diagonalizing Eq. (\ref{eq:multiequSch}), we can obtain the physical masses of the bare states after they are corrected by the hadron channels. If the off-diagonal terms in this equation vanish, the system reduces to two independent coupled-channel equations. The effective Hamiltonian in Eq.~(\ref{eq:multiequSch}) explicitly shows that the hadron-hadron channels not only provide mass corrections to the bare states but also induce additional mixing between the bare configurations and the hadron-hadron channels.

\section{Evaluation of matrix elements}\label{sec3}

In this section, we provide a detailed description of the calculation method for the relevant Hamiltonian in Eq.~(\ref{eq:multiequSch}). For each $J^P$ sector considered in this work, the required matrix elements include the diagonal bare state matrix elements $M_{\alpha_1}$ and $M_{\alpha_2}$, the off-diagonal bare state mixing term $M'$, and the transition amplitudes $H_{\alpha_i\to\beta_j}(\mathbf p)$ between the bare $\Omega_b$ cores and the nearby hadron channels. The bare sector matrix elements are calculated in a nonrelativistic constituent quark model, while the couplings to the hadron-hadron channels are calculated with the quark pair creation (QPC) model. These ingredients determine both the hadron-hadron channel corrections and the additional mixing effects discussed in Sec.~\ref{sec2}.

\subsection{Quark model description of the bare $\Omega_b$ cores}\label{sec:c/bsint}

We first determine the bare state matrix elements $M_{\alpha_1}$, $M_{\alpha_2}$, and $M'$. The bare $\Omega_b$ configurations are described in a nonrelativistic constituent quark model, and their masses and wave functions are obtained by solving the three-body Schr\"odinger equation. The resulting eigenfunctions are then used to evaluate the diagonal and off-diagonal matrix elements of the bare Hamiltonian $\hat H_0$. Following Refs.~\cite{Luo:2021dvj,Luo:2023sra,Luo:2023sne,Peng:2024pyl,Zhang:2025vtc}, we adopt
\begin{equation}\label{eq:H}
\hat{H}=\sum\limits_{i}\left(m_i+\frac{p_i^2}{2m_i}\right)+\sum\limits_{i<j}\left(H_{ij}^{\rm conf}+H_{ij}^{\rm hyp}+H_{ij}^{\rm so(cm)}+H_{ij}^{\rm so(tp)}\right),
\end{equation}
where $m_i$ and $p_i$ denote the mass and momentum of the $i$-th constituent quark. The interaction terms $H_{ij}^{\rm conf}$, $H_{ij}^{\rm hyp}$, $H_{ij}^{\rm so(cm)}$, and $H_{ij}^{\rm so(tp)}$ represent the spin-independent Cornell potential, the hyperfine interaction, the color-magnetic spin-orbit interaction, and the Thomas-precession spin-orbit interaction, respectively. Their explicit forms are
\begin{equation}\label{eq:Vconf}
H_{ij}^{\rm conf}=-\frac{2}{3}\frac{\alpha_s}{r_{ij}}+\frac{1}{2}br_{ij}+\frac{1}{2}C,
\end{equation}
\begin{equation}\label{eq:Vhyp}
\begin{split}
H_{ij}^{\rm hyp}=&\frac{2\alpha_s}{3m_im_j}\left[\frac{8\pi}{3}\tilde{\delta}(r_{ij}){\bf s}_i\cdot{\bf s}_j+\frac{1}{r_{ij}^3}S({\bf r},{\bf s}_i,{\bf s}_j)\right],
\end{split}
\end{equation}
\begin{equation}\label{eq:Vsocm}
\begin{split}
H_{ij}^{{\rm so(cm)}}=&\frac{2\alpha_s}{3r_{ij}^3}\left(\frac{{\bf r}_{ij}\times{\bf p}_i\cdot{\bf s}_i}{m_i^2}-\frac{{\bf r}_{ij}\times{\bf p}_j\cdot{\bf s}_j}{m_j^2}\right.\\
&\left.-\frac{{\bf r}_{ij}\times{\bf p}_j\cdot{\bf s}_i-{\bf r}_{ij}\times{\bf p}_i\cdot{\bf s}_j}{m_im_j}\right),
\end{split}
\end{equation}
\begin{equation}\label{eq:Vsotp}
H_{ij}^{{\rm so(tp)}}=-\frac{1}{2r_{ij}}\frac{\partial H_{ij}^{\rm conf}}{\partial r_{ij}}\left(\frac{{\bf r}_{ij}\times{\bf p}_i\cdot{\bf s}_i}{m_i^2}-\frac{{\bf r}_{ij}\times{\bf p}_j\cdot{\bf s}_j}{m_j^2}\right).
\end{equation}
Here, $\alpha_s$ is the one-gluon-exchange coupling constant, $b$ is the strength of
the linear confinement, and $C$ is the mass-renormalized constant. The smeared delta function is taken as
\begin{equation}
\tilde{\delta}(r)=\frac{\sigma^3}{\pi^{3/2}}{\rm e}^{-\sigma^2r^2},
\end{equation}
with the smearing parameter $\sigma$, and the tensor operator is defined by
\begin{equation}
S({\bf r},{\bf s}_i,{\bf s}_j)=\frac{3{\bf s}_i\cdot{\bf r}_{ij}{\bf s}_j\cdot{\bf r}_{ij}}{r_{ij}^2}-{\bf s}_i\cdot{\bf s}_j.
\end{equation}

For the singly bottom baryon $\Omega_b$, which is composed of two strange quarks and one bottom quark, we describe the three-body system in the $\rho$-$\lambda$ Jacobi coordinate system, as illustrated in Fig.~\ref{fig:jacodi}. The Jacobi coordinates are defined as
\begin{equation}
\boldsymbol{\rho}=\mathbf r_2-\mathbf r_1,\qquad
\boldsymbol{\lambda}=\mathbf r_3-\frac{m_1\mathbf r_1+m_2\mathbf r_2}{m_1+m_2},
\end{equation}
where particles 1 and 2 denote the two strange quarks and particle 3 denotes the bottom quark. For the $\Omega_b$ system with $m_1=m_2=m_s$, the $\rho$ coordinate describes the relative motion inside the $ss$ pair, while the $\lambda$ coordinate describes the motion of the bottom quark with respect to the center of mass of the $ss$ pair. Thus, the $\rho$ mode corresponds to the internal excitation of the two strange quarks, whereas the $\lambda$ mode represents the excitation between the strange-quark pair and the bottom quark.

The complete wave function of the $\Omega_b$ baryon can be written as
\begin{equation}
\Psi^{\Omega_b}_{JM_J}
=
\sum_{\alpha} C^{(\alpha)} \Psi^{\Omega_b,(\alpha)}_{JM_J}.
\end{equation}
Eq.~(\ref{eq:JMJ}) only contains the spin-spatial part, the complete basis can be written as
\begin{equation}
\Psi^{\Omega_b,(\alpha)}_{JM_J}
=
\chi^{\rm color}\,
\psi^{\rm flavor}_{\Omega_b}\,
\left[\left[
\chi_{s_{12}}^{\rm spin}\Phi^{\Omega_b,(\alpha)}_{L}(\boldsymbol{\rho},\boldsymbol{\lambda})\right]_{j_\ell}
\chi^{\rm spin}_{s_3}
\right]_{JM_J}.
\end{equation}
Here, $\alpha$ denotes the possible set of quantum numbers, including the orbital angular momenta, spin configurations, and quantum numbers of the spatial bases. The color wave function is the antisymmetric color-singlet one,
\begin{equation}
\chi^{\rm color}
=
\frac{1}{\sqrt{6}}
(rgb-rbg+gbr-grb+brg-bgr),
\end{equation}
and the flavor wave function of the $\Omega_b$ baryon is
\begin{equation}
\psi^{\rm flavor}_{\Omega_b}=s_1s_2b_3 .
\end{equation}
The $\chi_{s_{12}}^{\rm spin}$ and $\chi_{s_3}^{\rm spin}$ are spin wave functions of the first two light flavor quarks and the heavy flavor quark, respectively. Since the two strange quarks are identical, the total wave function should satisfy the Pauli principle under the exchange of the two strange quarks. With the antisymmetric color wave function and the symmetric $ss$ flavor component, the spin-spatial part is required to be symmetric under the exchange $1\leftrightarrow 2$.

Following the Gaussian expansion method adopted in this work, we expand the spatial wave function in terms of the Gaussian basis functions
\begin{equation}
\phi_{nlm}(\mathbf r)
=
N_{nl}\,r^l e^{-\nu_n r^2}Y_{lm}(\hat{\mathbf r}),
\end{equation}
where
\begin{equation}
\begin{split}
N_{nl} = \sqrt{\frac{2^{l+2}\left(2 v_n\right)^{l+\frac{3}{2}}}{\sqrt{\pi}(2 l+1)!!}},&\qquad\nu_n=\frac{1}{r_n^2},\\
r_n=r_{\min}a^{n-1},&\qquad
a=\left(\frac{r_{\max}}{r_{\min}}\right)^{1/(n_{\max}-1)},
\end{split}
\end{equation}
with $n=1 \cdots n_{\rm nmax}$. The spatial wave function is then expressed as
\begin{equation}
\Phi^{\Omega_b,(\alpha)}_{LM_L}(\boldsymbol{\rho},\boldsymbol{\lambda})
=
\left[
\phi_{n_\rho^g l_\rho}(\boldsymbol{\rho})
\phi_{n_\lambda^g l_\lambda}(\boldsymbol{\lambda})
\right]_{LM_L},
\end{equation}
where $l_\rho$ and $l_\lambda$ are the orbital angular momenta associated with the $\rho$- and $\lambda$-mode motions, respectively, and the bracket denotes the angular-momentum coupling to the total orbital angular momentum $L$. In summary, the label $\alpha$ collectively denotes the set
\begin{equation}
\begin{split}
\{s_{12}, n_\rho^g, l_\rho, n_\lambda^g, l_\lambda, L, j_\ell \}.
\end{split}
\end{equation}
It should be emphasized that $n_\rho^g$ and $n_\lambda^g$ are Gaussian-basis indices rather than the radial quantum numbers of the baryon. The wave functions corresponding to different radial quantum numbers $n_\rho$ and $n_\lambda$ are obtained through the expansion coefficients $C^{(\alpha)}$, with fixed $\{s_{12},l_\rho,l_\lambda,L,j_\ell\}$ and different Gaussian-basis indices $n_\rho^g$ and $n_\lambda^g$. The coefficients $C^{(\alpha)}$ are determined using the Rayleigh--Ritz variational method, whose details are presented below.

With the above basis, the kinetic-energy, potential-energy, and overlap matrix elements are evaluated as
\begin{equation}
\begin{split}
T^{\alpha'\alpha}
=
\left\langle
\Psi^{\Omega_b,(\alpha')}_{JM_J}
\left|T\right|
\Psi^{\Omega_b,(\alpha)}_{JM_J}
\right\rangle ,\\
V^{\alpha'\alpha}
=
\left\langle
\Psi^{\Omega_b,(\alpha')}_{JM_J}
\left|V\right|
\Psi^{\Omega_b,(\alpha)}_{JM_J}
\right\rangle ,
\end{split}
\end{equation}
and
\begin{equation}
N^{\alpha'\alpha}
=
\left\langle
\Psi^{\Omega_b,(\alpha')}_{JM_J}
\middle|
\Psi^{\Omega_b,(\alpha)}_{JM_J}
\right\rangle .
\end{equation}
The three-body Schrödinger equation is then solved by the Rayleigh--Ritz variational method \cite{Hiyama:2003cu,Luo:2023sne,Peng:2024pyl,Zhang:2025vtc}, which leads to the generalized eigenvalue equation
\begin{equation}
\sum_{\alpha}
\left(
T^{\alpha'\alpha}+V^{\alpha'\alpha}
\right)
C^{(\alpha)}
=
E
\sum_{\alpha}
N^{\alpha'\alpha}
C^{(\alpha)} . 
\end{equation}
By diagonalizing this equation, we obtain the bare mass spectrum of the $\Omega_b$ states together with the corresponding expansion coefficients. These eigenfunctions are further used to evaluate the bare-state matrix elements entering Eq.~(\ref{eq:multiequSch}), namely
\begin{equation}
M_{\alpha_i}
=
\left\langle
\Psi^{\Omega_b}_{\alpha_i}
\left|H_{3q}\right|
\Psi^{\Omega_b}_{\alpha_i}
\right\rangle ,
\qquad
M'
=
\left\langle
\Psi^{\Omega_b}_{\alpha_1}
\left|H_{3q}\right|
\Psi^{\Omega_b}_{\alpha_2}
\right\rangle ,
\end{equation}
where $H_{3q}=T+V$ is the three-quark Hamiltonian.

To carry out the numerical calculation, we next need to determine the parameters of the conventional potential model. Since only the ground state $\Omega_b(1S)$ has been experimentally well established so far, the $\Omega_b$ spectrum alone cannot provide sufficient constraints to determine the potential model parameters. By invoking heavy-quark symmetry, we adopt the parameters $\alpha_s$, $b$, and $\sigma$ from the $\Omega_c$ sector, while the renormalization constant $C$ is fixed by reproducing the mass of the $\Omega_b(1S)$ state \cite{Luo:2023sne,Zhang:2025vtc}. In this way, the parameters of the conventional potential model are fixed as $m_n=0.370$ GeV, $m_s=0.600$ GeV, $m_b=5.210$ GeV, $\alpha_s=0.578$, $b=0.144$, $\sigma=1.732~\mathrm{GeV}$, and $C=-0.646~\mathrm{GeV}$.

\subsection{Coupling of the bare $\Omega_b$ cores to the hadron-hadron channels}\label{sec:XiKbareCoup}

We now turn to the transition matrix elements $H_{\alpha_i\to\beta_j}(\mathbf p)$ defined in Eq.~(\ref{eq:TrMChannel}), which couple the bare $\Omega_b$ configurations to the nearby hadron channels. In the present work, these transition amplitudes are evaluated within the quark-pair-creation (QPC) model~\cite{Micu:1968mk,LeYaouanc:1972vsx}, for which the interaction Hamiltonian is given by
\begin{equation}
\hat{H}_I=g\int \textrm{d}^3x\,\bar{\psi}(x)\psi(x).
\end{equation}
Here $g=2m_q\gamma$, where $m_q$ denotes the mass of the created light quark and $\gamma$ is the dimensionless quark-anquark pair creation strength from the vacuum.

In the nonrelativistic limit, the QPC operator takes the form~\cite{Chen:2016iyi}
\begin{equation}
\begin{split}\label{eq:3p0gamma}
\hat{H}_I=&-3\gamma\sum_{m}\left<1,m;1,-m|0,0\right>\int \textrm{d}^3\textbf{p}_{\mu}d^3\textbf{p}_{\nu}\delta\left(\textbf{p}_{\mu}+\textbf{p}_{\nu}\right)\\
&\times\mathcal{Y}_1^m\left(\frac{\textbf{p}_{\mu}-\textbf{p}_{\nu}}{2}\right)\omega^{\left(\mu,\nu\right)}\phi^{\left(\mu,\nu\right)}\chi_{-m}^{\left(\mu,\nu\right)}b_{\mu}^{\dagger}\left(\textbf{p}_{\mu}\right)d_{\nu}^{\dagger}\left(\textbf{p}_{\nu}\right)\, ,
\end{split}
\end{equation}
where $\omega$, $\phi$, $\chi$, and $\mathcal{Y}_1^m$ denote the color, flavor, spin, and orbital parts of the created $q\bar q$ pair from the vacuum, respectively. The creation operators of quark and antiquark are written as $b_{\mu}^{\dagger}$ and $d_{\nu}^{\dagger}$.

The determination of the dimensionless parameter $\gamma$ in Eq.~(\ref{eq:3p0gamma}) is essential for evaluating the partial wave amplitude of the singly bottom baryons sector. In this work, we fix $\gamma=9.61$ by reproducing the experimental width of the $\Xi_b^{*}(5835)$ state \cite{ParticleDataGroup:2026mpi}. The resulting value is then adopted in all subsequent calculations.

In analogy with the treatment adopted in Ref.~\cite{Luo:2021dvj,Chen:2016iyi,Luo:2023sne,Luo:2023sra,Peng:2024pyl}, we adopt the $\Omega_b$ spatial wave function appearing in the strong-decay calculation by a simple harmonic oscillator form. The parameters $\beta_\rho$ and $\beta_\lambda$ of the simple harmonic oscillator wave function are determined from the corresponding root-mean-square radii extracted from the numerical GEM wave function, so that the dominant size effect of the three-quark configuration is preserved. This method greatly simplifies the calculation while maintaining the essential features of the baryon spatial structure. Following this calculation procedure, the values of $\beta$ used in this work are listed in Table~\ref{table:betaV}.

\begin{table*}\label{table:betaV}
\centering
\caption{The $\beta$ parameters of the hadrons relevant to this work in units of GeV.}
\renewcommand\arraystretch{1.25}
\begin{tabular*}{178mm}{@{\extracolsep{\fill}}ccccccccc}
\toprule[1.00pt]
\toprule[1.00pt]
States&$\beta_\rho$ &$\beta_\lambda$ &States &$\beta_\rho$ &$\beta_\lambda$&States&$\beta_\rho$ &$\beta_\lambda$\\
\midrule[0.75pt]
$\Omega_b(1S)$  &0.289 &0.455 &$\Xi_b(1S)$&0.297 &0.412&$\Xi_b^{*}(1S)$&0.241&0.389\\
$\Omega_b^*(1S)$  &0.283 &0.440&$\Xi_b(1P)$&0.272 &0.278&$\Xi_b^{\prime}(1P)$&0.223&0.276\\
$\Omega_b(1P)$  &0.260 &0.304 &$\Xi_b^{\prime}(1S)$&0.245 &0.398&\multicolumn{3}{c}{$\beta_K=0.385$}\\
\bottomrule[1.00pt]
\bottomrule[1.00pt]
\end{tabular*}
\end{table*}

Using the bare $\Omega_b$ wave functions described above, together with the wave functions of the two-hadron channels, we calculate the momentum-dependent amplitudes $H_{\alpha_i\to\beta_j}(\mathbf p)$. These amplitudes enter the self-energy terms $\Delta M$ and the hadron-hadron channel induced mixing terms defined in Sec.~\ref{sec2}, and therefore determine how the nearby thresholds modify the masses and compositions of the physical $\Omega_b(1P)$ states. In particular, they provide the key dynamical input for assessing whether the missing $J^P=1/2^-$ state can be significantly shifted toward, or even below, the $\Xi_b\bar K$ threshold.

\section{Results and discussions}\label{sec4}
Based on the unquenched quark model introduced in Sec.~\ref{sec2} and the matrix elements calculated in Sec.~\ref{sec3}, we can now explore the unquenched effects in the $\Omega_b(1P)$ sector. In this section, we present our numerical results and discuss their implications for the $\Omega_b(1P)$ spectrum. We will show how the inclusion of coupled-channel effects leads to mass shifts and configuration mixing, and how these modifications affect the interpretation of the physical states. These results are expected to provide further insight into the internal structure of the observed $\Omega_b$ resonances.

With these potential model parameters in Sec.~\ref{sec3}, the bare masses are obtained as $M_{\Omega_{b0}^d(1P,1/2^-)}=6354$ MeV and $M_{\Omega_{b1}^d(1P,1/2^-)}=6344$ MeV\footnote{In the notation $\Omega_{b j_\ell}^d$, the subscript specifies the total angular momentum $j_\ell$ of the light degrees of freedom, and the superscript $d$ denotes a physical state dominated by the corresponding $j_\ell$ configuration. Accordingly, $\Omega_{b0}^d$ and $\Omega_{b1}^d$ refer to the states dominated by the $j_\ell=0$ and $j_\ell=1$ configurations, respectively.}, with an off-diagonal mixing term of $5$ MeV. For the $J^P=3/2^-$ sector, the two bare state masses are found to be $6363$ and $6360$ MeV, respectively, with a corresponding mixing term of $1$ MeV. Overall, these $1P$ $\Omega_b$ bare states are clustered around $6.35$~GeV, with only small mass splittings and off-diagonal mixings of a few MeV. This indicates that the mixing among different basis configurations remains weak at the bare state level, so that the bare state basis still provides a reasonable starting point for discussing the spectrum. However, because these states are close in mass, the mass shifts and configuration renormalization induced by hadron loops may play an important role in determining the final level ordering and compositions of the physical states.

\begin{figure}
    \centering
    \includegraphics[width=0.48\textwidth]{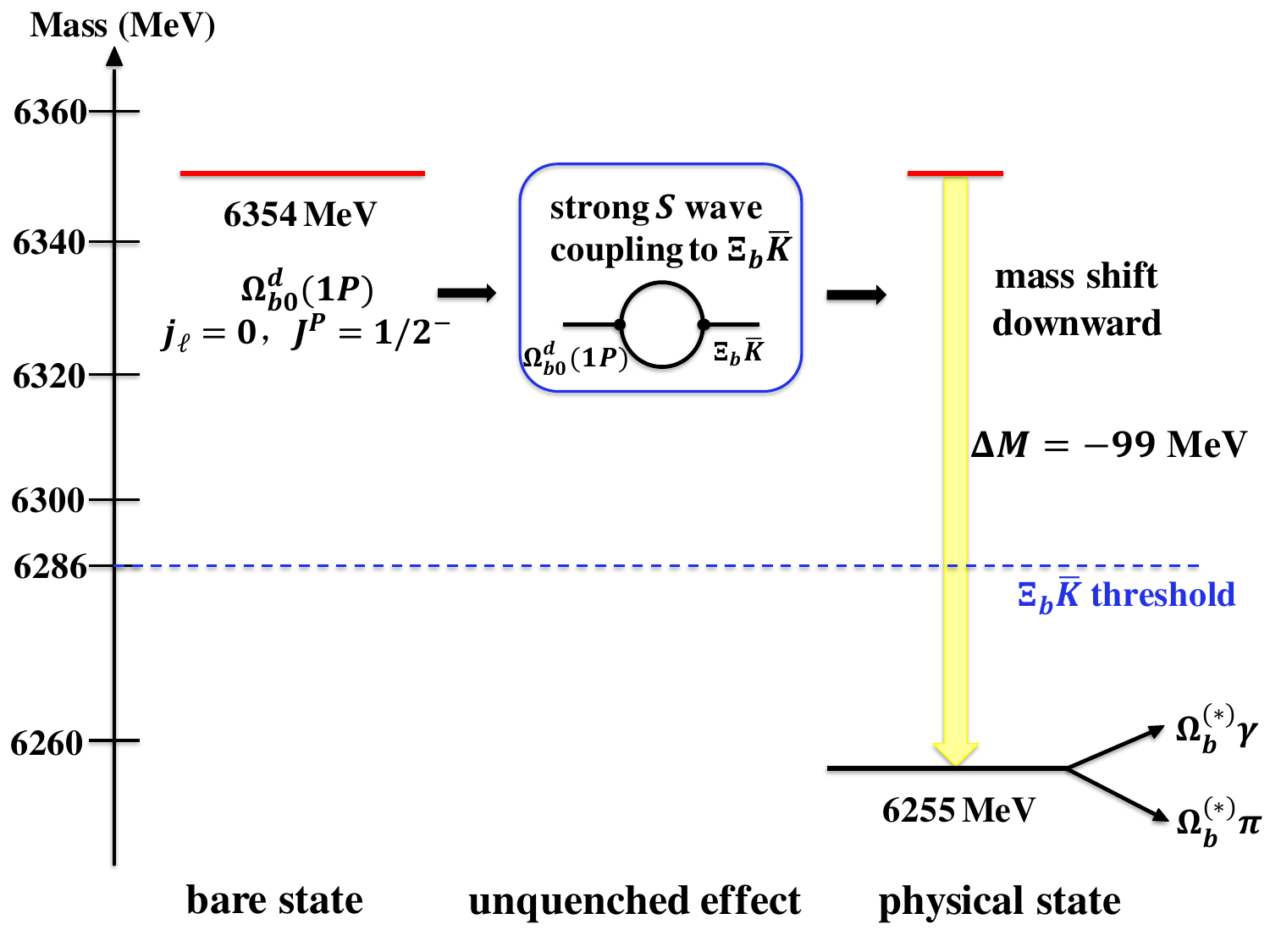}
    \caption{Bare-to-physical evolution of the $\Omega_{b0}^d(1P)$ state under $\Xi_b\bar{K}$ coupling and its possible observable channels.}
    \label{fig:omegaj0}
\end{figure}

After incorporating the unquenched effects, the masses of $\Omega_{b0}^d(1P,1/2^-)$ and $\Omega_{b1}^d(1P,1/2^-)$ are lowered to $6255$~MeV and $6313$ MeV, respectively. Similarly, the resulting masses of $\Omega_{b1}^d(1P,3/2^-)$ and $\Omega_{b2}^d(1P,3/2^-)$ are $6328$ MeV and $6362$~MeV, respectively. We now turn to a detailed discussion of these states.

Among these states, the most remarkable one is $\Omega_{b0}^d(1P,1/2^-)$. Its mass is shifted downward from the bare value $6.354~\mathrm{GeV}$ to $6.255~\mathrm{GeV}$, corresponding to a sizable downward shift of about $99~\mathrm{MeV}$. This process is schematically illustrated in the Fig.~\ref{fig:omegaj0}. Such a large renormalization implies that this state is strongly influenced by the nearby hadron-hadron channel. 

The component analysis in Eq.~(\ref{eq:Norm}) further supports this interpretation. For $\Omega_{b0}^d(1P,1/2^-)$, one obtains
\[
P(\Omega_{b0})=49.6\%,\qquad P(\Omega_{b1})=0.3\%,
\]
\[
P(\Xi_b\bar K)=50.0\%,\qquad P(\Xi_b^\prime \bar K)=0.1\%.
\]
The qualitative conclusion is clear: the physical state is dominated by the bare $\Omega_{b0}$ core and the $\Xi_b\bar K$ channel, while the admixtures from the $\Omega_{b1}$ configuration and the $\Xi_b^\prime \bar K$ channel are negligible. 

By diagonalizing the $\Omega_{b0}$–$\Omega_{b1}$ mixing matrix, we obtain a mixing angle of $\theta_{\Omega_{c0}^d(1P,1/2^-)}=-4.0^\circ$, which is particularly small. Such a weak mixing is sufficient to justify identifying the physical state as $\Omega_{b0}^d(1P,1/2^-)$. This state therefore remains dominated by the $\Omega_{b0}$ component, making the notation $\Omega_{b0}^d(1P,1/2^-)$ appropriate. In this sense, the large mass shift does not originate from a strong mixing between the two bare $1/2^-$ configurations themselves, but rather from the strong coupling of the $\Omega_{b0}$ configuration to the relevant hadron-hadron channel. The Ref.~\cite{Luo:2021dvj} gives a mixing angle of $-12.9^\circ$ for the $\Omega_c$ states, noticeably larger than the value in the $\Omega_b$ sector. This directly suggests that increasing the heavy-quark mass in singly heavy baryons leads to a more pronounced manifestation of heavy-quark symmetry.

A particularly important consequence of the unquenched effect is that the physical mass of $\Omega_{b0}^d(1P,1/2^-)$ is driven below the $\Xi_b\bar K$ threshold. Once this happens, the OZI-allowed strong decay $\Omega_{b0}^d(1P,1/2^-)\to \Xi_b\bar K$ becomes kinematically forbidden. Therefore, although the $\Xi_b\bar K$ component remains substantial in its wave function, this state cannot appear as an ordinary above-threshold resonance in the $\Xi_b\bar K$ invariant-mass spectrum. In other words, one should not expect to observe $\Omega_{b0}^d(1P,1/2^-)$ as a visible enhancement in the $\Xi_b\bar K$ channel. Instead, the closure of its dominant strong-decay mode implies that it should be a narrow state. In this sense, $\Omega_{b0}^d(1P,1/2^-)$ may be viewed as the bottom-strange counterpart of a threshold-driven low-mass state.

Of course, the subthreshold interpretation would make the state $\Omega_{b0}^d(1P,1/2^-)$ even more difficult to detect. This feature has an immediate implication for the experiments. Since the $\Xi_b\bar K$ mode is closed, it is not the most suitable channel in which to search for $\Omega_{b0}^d(1P,1/2^-)$. More promising discovery modes are expected to be radiative transitions to lower-lying $\Omega_b$ baryons, such as $\Omega_b\gamma$ and $\Omega_b^\ast\gamma$. In addition, isospin-violating hadron transitions, e.g. $\Omega_b^{(*)}\pi^0$, may also provide possible search channels, although such processes are expected to be suppressed. The latter can proceed only via the $\eta$-$\pi$ mixing mechanism and is therefore suppressed at the level of $\mathcal{O}(10^{-4})$. From an experimental point of view, a more realistic strategy would therefore be to perform dedicated searches for a narrow structure below the $\Xi_b\bar K$ threshold in the $\Omega_b^{(*)}\gamma$ and $\Omega_b^{(*)}\pi^0$ process, rather than in the $\Xi_b\bar K$ invariant-mass distribution.

Experimental precedents also support this search strategy. Radiative transitions have proven useful in studies of several single-heavy baryons \cite{CLEO:1998wvk,BaBar:2006pve,Solovieva:2008fw,Belle:2020ozq}, while the subthreshold $D_{s0}^\ast(2317)$ was discovered through its isospin-violating $D_s\pi^0$ process \cite{BaBar:2003oey}. More recently, the radiative decay $D_{s0}^\ast(2317)^+\to D_s^{\ast+}\gamma$ has been observed for the first time with a significance exceeding $10\sigma$ \cite{Belle:2025dtr}. These experimental advances demonstrate the increasing sensitivity to such suppressed decay modes and provide encouraging prospects for testing our prediction for the $\Omega_{b0}^d(1P,1/2^-)$ state.

By contrast, the other three $\Omega_b(1P)$ states behave more like conventional orbitally excited baryons with finite unquenched corrections. The $\Omega_{b1}^d(1P,1/2^-)$ and $\Omega_{b1}^d(1P,3/2^-)$ states are shifted downward by only $31$ and $35~\mathrm{MeV}$, respectively, while the coupled-channel effect on $\Omega_{b2}^d(1P,3/2^-)$ is almost negligible. Hence, the present results show that the coupled-channel effect does not modify the whole $\Omega_b(1P)$ multiplet uniformly. Rather, its impact depends strongly on the quantum numbers of each state and on the strength of its coupling to the nearby hadron channels.

The quantum-number analysis of the couplings between the baryon states and the hadron channels shows that the transitions $\Omega_{b0}(1P,1/2^-)\to \Xi_b\bar{K}$, $\Omega_{b1}(1P,1/2^-)\to \Xi_b^{\prime}\bar{K}$, and $\Omega_{b1}(1P,3/2^-)\to \Xi_b^{*}\bar{K}$ proceed through $S$-wave couplings, whereas the remaining amplitudes are governed by $D$-wave or higher partial waves. Accordingly, unquenched effects are expected to be more pronounced in the $S$-wave channels and less significant in the higher-partial-wave channels. This expectation is borne out by the mass shifts obtained in our calculation.

Moreover, by solving the coupled-channel equations, the mixing angles of the remaining three states are obtained as $\theta_{\Omega_{c1}^d(1P,1/2^-)}=-3.1^\circ$, $\theta_{\Omega_{c1}^d(1P,3/2^-)}=1.6^\circ$, and $\theta_{\Omega_{c2}^d(1P,3/2^-)}=1.3^\circ$, respectively. These values are much smaller than those predicted for the corresponding $\Omega_c$ states in the literature~\cite{Luo:2021dvj}, further indicating that heavy-quark symmetry is more manifest in the heavier singly bottom baryons.

Overall, the present analysis suggests that the coupled-channel effect plays a crucial role in shaping the $\Omega_b(1P)$ spectrum. Its most striking consequence is the sizable downward shift of $\Omega_{b0}^d(1P,1/2^-)$, which drives this state below the $\Xi_b\bar K$ threshold and makes it inaccessible in the $\Xi_b\bar K$ channel. This state should therefore be searched for as a narrow subthreshold structure in radiative or suppressed isospin-breaking decay channels, while the remaining $\Omega_b(1P)$ states can still be understood, to a good approximation, as conventional $1P$ excitations with moderate hadron-hadron channel corrections.

\section{Summary}\label{sec5}

 In this work, we have studied the low-lying $\Omega_b(1P)$ states within an unquenched coupled-channel framework. Our results show that the coupled-channel effect plays a strongly state-dependent role in the $\Omega_b(1P)$ spectrum. Among the four low-lying $1P$ states predicted in our model, the most significant mass renormalization occurs for the state dominated by the $j_\ell=0$ configuration with $J^P=\frac{1}{2}^-$, $\Omega_{b0}^d(1P,1/2^-)$, whose mass is shifted from the bare value $6.354~\mathrm{GeV}$ down to $6.255~\mathrm{GeV}$. By contrast, the other three $1P$ states receive only moderate or negligible corrections and remain close to their bare quark model masses.

 The component analysis indicates that $\Omega_{b0}^d(1P,1/2^-)$ develops a sizable $\Xi_b\bar K$ component through coupled-channel mixing, suggesting that this state cannot be regarded as a purely quenched three-quark excitation but rather as a strongly dressed state generated by its coupling to the nearby $S$-wave hadron channel. More importantly, after unquenching effects are included, the physical mass of $\Omega_{b0}^d(1P,1/2^-)$ is shifted below the $\Xi_b\bar K$ threshold. Consequently, its OZI-allowed strong decay into $\Xi_b\bar K$ becomes kinematically forbidden, preventing it from appearing as a conventional $\Xi_b^0K^-$ resonance above threshold. This may provide a possible explanation for why a low-lying $J^P=\frac{1}{2}^-$ $\Omega_b$ state has not yet been identified experimentally.

Our results highlight the importance of unquenched dynamics in the $\Omega_b(1P)$ sector. Observing the predicted subthreshold $\Omega_{b0}^d(1P,1/2^-)$ state experimentally would provide valuable evidence supporting threshold-induced mass renormalization effects in bottom baryons and would offer an instructive comparison with similar phenomena in the charmed sector. Compared with the much more developed spectroscopy of the $\Omega_c(1P)$ family, experimental information on excited $\Omega_b$ baryons remains limited, and the low-lying $J^P=\frac{1}{2}^-$ state represents an important target for future searches.

Because the strong $\Xi_b\bar K$ decay channel is closed, the most favorable discovery modes are expected to be radiative transitions, such as $\Omega_b\gamma$ and $\Omega_b^\ast\gamma$, together with suppressed isospin-breaking decays such as $\Omega_b^{(*)}\pi^0$. With heavy-flavor spectroscopy entering a precision era, the rapidly increasing Run-3 data samples and improving experimental sensitivity at LHCb make dedicated searches for such rare radiative and isospin-breaking decay modes increasingly feasible, providing a realistic opportunity to test our theoretical prediction.

\section*{ACKNOWLEDGMENTS}
This work is supported by the Natural Science Foundation of Gansu Province (No. 26RCKA012 and No. 25JRRA799), the National Natural Science Foundation of China under Grants No. 12405098, No. 12335001, No. 1267050014, No. 12275067, and No. 12247101, Science and Technology Innovation Leading Talent Support Program of Henan Province (Grant No. 254200510039), Science and Technology R$\&$D Program Joint Fund Project of Henan Province  (Grant No.225200810030), Henan Province Foreign Scientists Studio (Grant No. GZS2020032), National Key R$\&$D Program of China (Grant No. 2023YFA1606000), the Natural Science Foundation of Henan Province (Grant No. 252300421999), the ‘111 Center’ under Grant No. B20063, the fundamental Research Funds for the Central Universities, and the Talent Scientific Fund of Lanzhou University.

\bibliographystyle{UserDefined}
\bibliography{References}

\begin{thebibliography}{10}

\bibitem{Chen:2016spr}
H.~X. Chen, W.~Chen, X.~Liu, Y.~R. Liu, and S.~L. Zhu,
A review of the open charm and open bottom systems,
\href{https://doi.org/10.1088/1361-6633/aa6420}{Rept. Prog. Phys. {\bf 80},
  076201 (2017)}.

\bibitem{Guo:2017jvc}
F.~K. Guo, C.~Hanhart, U.~G. Mei{\ss}ner, Q.~Wang, Q.~Zhao, and B.~S. Zou,
Hadronic molecules,
\href{https://doi.org/10.1103/RevModPhys.90.015004}{Rev. Mod. Phys. {\bf 90},
  015004 (2018)},
[Erratum: Rev.Mod.Phys. 94, 029901 (2022)].

\bibitem{Cheng:2015iom}
H.~Y. Cheng,
Charmed baryons circa 2015,
\href{https://doi.org/10.1007/s11467-015-0483-z}{Front. Phys. (Beijing) {\bf
  10}, 101406 (2015)}.

\bibitem{Brambilla:2019esw}
N.~Brambilla, S.~Eidelman, C.~Hanhart, A.~Nefediev, C.~P. Shen, C.~E. Thomas,
  A.~Vairo, and C.~Z. Yuan,
The $XYZ$ states: experimental and theoretical status and perspectives,
\href{https://doi.org/10.1016/j.physrep.2020.05.001}{Phys. Rept. {\bf 873},
  1--154 (2020)}.

\bibitem{Liu:2019zoy}
Y.~R. Liu, H.~X. Chen, W.~Chen, X.~Liu, and S.~L. Zhu,
Pentaquark and Tetraquark states,
\href{https://doi.org/10.1016/j.ppnp.2019.04.003}{Prog. Part. Nucl. Phys. {\bf
  107}, 237--320 (2019)}.

\bibitem{Chen:2022asf}
H.~X. Chen, W.~Chen, X.~Liu, Y.~R. Liu, and S.~L. Zhu,
An updated review of the new hadron states,
\href{https://doi.org/10.1088/1361-6633/aca3b6}{Rept. Prog. Phys. {\bf 86},
  026201 (2023)}.

\bibitem{Dong:2021juy}
X.~K. Dong, F.~K. Guo, and B.~S. Zou,
A survey of heavy-antiheavy hadronic molecules,
\href{https://doi.org/10.13725/j.cnki.pip.2021.02.001}{Progr. Phys. {\bf 41},
  65--93 (2021)}.

\bibitem{Cheng:2021qpd}
H.~Y. Cheng,
Charmed baryon physics circa 2021,
\href{https://doi.org/10.1016/j.cjph.2022.06.021}{Chin. J. Phys. {\bf 78},
  324--362 (2022)}.

\bibitem{Bai:2026atm}
Z.~Y. Bai, D.~Y. Chen, Q.~Huang, X.~Liu, S.~Q. Luo, and J.~Z. Wang,
Unquenched charmonium and beyond,
\href{https://doi.org/10.1016/j.physrep.2026.08.002}{Phys. Rept. {\bf 1204},
  1--162 (2027)}.

\bibitem{Wang:2025dur}
X.~Wang, X.~Liu, and Y.~Gao,
Colloquium: Hadron production in open-charm meson pairs at $e^+e^-$ colliders,
\href{https://doi.org/10.1103/2mrp-chly}{Rev. Mod. Phys. {\bf 98}, 021001
  (2026)}.

\bibitem{Liu:2024uxn}
M.~Z. Liu, Y.~W. Pan, Z.~W. Liu, T.~W. Wu, J.~X. Lu, and L.~S. Geng,
Three ways to decipher the nature of exotic hadrons: Multiplets, three-body
  hadronic molecules, and correlation functions,
\href{https://doi.org/10.1016/j.physrep.2024.12.001}{Phys. Rept. {\bf 1108},
  1--108 (2025)}.

\bibitem{Meng:2022ozq}
L.~Meng, B.~Wang, G.~J. Wang, and S.~L. Zhu,
Chiral perturbation theory for heavy hadrons and chiral effective field theory
  for heavy hadronic molecules,
\href{https://doi.org/10.1016/j.physrep.2023.04.003}{Phys. Rept. {\bf 1019},
  1--149 (2023)}.

\bibitem{LHCb:2020tqd}
R.~Aaij {\it et~al}. (LHCb Collaboration),
First observation of excited $\Omega_b^-$ states,
\href{https://doi.org/10.1103/PhysRevLett.124.082002}{Phys. Rev. Lett. {\bf
  124}, 082002 (2020)}.

\bibitem{Wang:2017kfr}
K.~L. Wang, Y.~X. Yao, X.~H. Zhong, and Q.~Zhao,
Strong and radiative decays of the low-lying $S$- and $P$-wave singly heavy
  baryons,
\href{https://doi.org/10.1103/PhysRevD.96.116016}{Phys. Rev. D {\bf 96}, 116016
  (2017)}.

\bibitem{Xiao:2020oif}
L.~Y. Xiao, K.~L. Wang, M.~S. Liu, and X.~H. Zhong,
Possible interpretation of the newly observed $\Omega_b$ states,
\href{https://doi.org/10.1140/epjc/s10052-020-7823-z}{Eur. Phys. J. C {\bf 80},
  279 (2020)}.

\bibitem{Liang:2020hbo}
W.~Liang and Q.~F. L{\"u},
Strong decays of the newly observed narrow $\Omega_b$ structures,
\href{https://doi.org/10.1140/epjc/s10052-020-7759-3}{Eur. Phys. J. C {\bf 80},
  198 (2020)}.

\bibitem{Yang:2020zrh}
H.~M. Yang and H.~X. Chen,
$P$-wave bottom baryons of the $SU(3)$ flavor $\mathbf{6}_F$,
\href{https://doi.org/10.1103/PhysRevD.101.114013}{Phys. Rev. D {\bf 101},
  114013 (2020)},
[Erratum: Phys.Rev.D 102, 079901 (2020)].

\bibitem{Karliner:2020fqe}
M.~Karliner and J.~L. Rosner,
Interpretation of excited $\Omega_b$ signals,
\href{https://doi.org/10.1103/PhysRevD.102.014027}{Phys. Rev. D {\bf 102},
  014027 (2020)}.

\bibitem{Chen:2020mpy}
H.~X. Chen, E.~L. Cui, A.~Hosaka, Q.~Mao, and H.~M. Yang,
Excited $\Omega_b$ baryons and fine structure of strong interaction,
\href{https://doi.org/10.1140/epjc/s10052-020-7824-y}{Eur. Phys. J. C {\bf 80},
  256 (2020)}.

\bibitem{Wang:2020pri}
Z.~G. Wang,
Analysis of the $\Omega_b(6316)$, $\Omega_b(6330)$, $\Omega_b(6340)$ and
  $\Omega_b(6350)$ with QCD sum rules,
\href{https://doi.org/10.1142/S0217751X20500438}{Int. J. Mod. Phys. A {\bf 35},
  2050043 (2020)}.

\bibitem{XuYongJiang:2020cht}
Y.~L.~L. Yong-Jiang~Xu and M.~Q. Huang,
$P$-wave $\Omega_b$ states: masses and pole residues,
\href{https://doi.org/10.1088/1674-1137/ac3df2}{Chin. Phys. C {\bf 46}, 043103
  (2022)}.

\bibitem{Mao:2015gya}
Q.~Mao, H.~X. Chen, W.~Chen, A.~Hosaka, X.~Liu, and S.~L. Zhu,
QCD sum rule calculation for $P$-wave bottom baryons,
\href{https://doi.org/10.1103/PhysRevD.92.114007}{Phys. Rev. D {\bf 92}, 114007
  (2015)}.

\bibitem{Mutuk:2020rzm}
H.~Mutuk,
A study of excited $\Omega _b^-$ states in hypercentral constituent quark model
  via artificial neural network,
\href{https://doi.org/10.1140/epja/s10050-020-00161-5}{Eur. Phys. J. A {\bf
  56}, 146 (2020)}.

\bibitem{Luo:2024jov}
X.~Luo, H.~M. Yang, and H.~X. Chen,
Radiative decays of $P$-wave bottom baryons from light-cone sum rules,
\href{https://doi.org/10.1103/PhysRevD.111.056027}{Phys. Rev. D {\bf 111},
  056027 (2025)}.

\bibitem{Liang:2017ejq}
W.~H. Liang, J.~M. Dias, V.~R. Debastiani, and E.~Oset,
Molecular $\Omega_b$ states,
\href{https://doi.org/10.1016/j.nuclphysb.2018.03.008}{Nucl. Phys. B {\bf 930},
  524--532 (2018)}.

\bibitem{Liang:2020dxr}
W.~H. Liang and E.~Oset,
Observed $\Omega_b$ spectrum and meson-baryon molecular states,
\href{https://doi.org/10.1103/PhysRevD.101.054033}{Phys. Rev. D {\bf 101},
  054033 (2020)}.

\bibitem{ParticleDataGroup:2026mpi}
F.~Takahashi {\it et~al}. (Particle Data Group),
Review of Particle Physics,
\href{https://doi.org/10.1142/s0217751x26300115}{Int. J. Mod. Phys. A {\bf 41},
  2630011 (2027)}.

\bibitem{LHCb:2017uwr}
R.~Aaij {\it et~al}. (LHCb Collaboration),
Observation of five new narrow $\Omega_c^0$ states decaying to $\Xi_c^+ K^-$,
\href{https://doi.org/10.1103/PhysRevLett.118.182001}{Phys. Rev. Lett. {\bf
  118}, 182001 (2017)}.

\bibitem{LHCb:2021ptx}
R.~Aaij {\it et~al}. (LHCb Collaboration),
Observation of excited $\Omega_c^0$ baryons in $\Omega_b^- \to \Xi_c^+
  K^-\pi^-$decays,
\href{https://doi.org/10.1103/PhysRevD.104.L091102}{Phys. Rev. D {\bf 104},
  L091102 (2021)}.

\bibitem{LHCb:2023sxp}
R.~Aaij {\it et~al}. (LHCb Collaboration),
Observation of New $\Omega_c^0$ States Decaying to the $\Xi_c^+K^-$ Final
  State,
\href{https://doi.org/10.1103/PhysRevLett.131.131902}{Phys. Rev. Lett. {\bf
  131}, 131902 (2023)}.

\bibitem{Belle:2017ext}
J.~Yelton {\it et~al}. (Belle Collaboration),
Observation of Excited $\Omega_c$ Charmed Baryons in $e^+e^-$ Collisions,
\href{https://doi.org/10.1103/PhysRevD.97.051102}{Phys. Rev. D {\bf 97}, 051102
  (2018)}.

\bibitem{Chen:2017sci}
H.~X. Chen, Q.~Mao, W.~Chen, A.~Hosaka, X.~Liu, and S.~L. Zhu,
Decay properties of $P$-wave charmed baryons from light-cone QCD sum rules,
\href{https://doi.org/10.1103/PhysRevD.95.094008}{Phys. Rev. D {\bf 95}, 094008
  (2017)}.

\bibitem{Karliner:2017kfm}
M.~Karliner and J.~L. Rosner,
Very narrow excited $\Omega_c$ baryons,
\href{https://doi.org/10.1103/PhysRevD.95.114012}{Phys. Rev. D {\bf 95}, 114012
  (2017)}.

\bibitem{Wang:2017hej}
K.~L. Wang, L.~Y. Xiao, X.~H. Zhong, and Q.~Zhao,
Understanding the newly observed $\Omega_c$ states through their decays,
\href{https://doi.org/10.1103/PhysRevD.95.116010}{Phys. Rev. D {\bf 95}, 116010
  (2017)}.

\bibitem{Wang:2017vnc}
W.~Wang and R.~L. Zhu,
Interpretation of the newly observed $\Omega_c^0$ resonances,
\href{https://doi.org/10.1103/PhysRevD.96.014024}{Phys. Rev. D {\bf 96}, 014024
  (2017)}.

\bibitem{Padmanath:2017lng}
M.~Padmanath and N.~Mathur,
Quantum Numbers of Recently Discovered $\Omega^{0}_{c}$ Baryons from Lattice
  QCD,
\href{https://doi.org/10.1103/PhysRevLett.119.042001}{Phys. Rev. Lett. {\bf
  119}, 042001 (2017)}.

\bibitem{Cheng:2017ove}
H.~Y. Cheng and C.~W. Chiang,
Quantum numbers of $\Omega_c$ states and other charmed baryons,
\href{https://doi.org/10.1103/PhysRevD.95.094018}{Phys. Rev. D {\bf 95}, 094018
  (2017)}.

\bibitem{Wang:2017zjw}
Z.~G. Wang,
Analysis of $\Omega _c(3000)$ , $\Omega _c(3050)$ , $\Omega _c(3066)$ , $\Omega
  _c(3090)$ and $\Omega _c(3119)$ with QCD sum rules,
\href{https://doi.org/10.1140/epjc/s10052-017-4895-5}{Eur. Phys. J. C {\bf 77},
  325 (2017)}.

\bibitem{Zhao:2017fov}
Z.~Zhao, D.~D. Ye, and A.~Zhang,
Hadronic decay properties of newly observed $\Omega_c$ baryons,
\href{https://doi.org/10.1103/PhysRevD.95.114024}{Phys. Rev. D {\bf 95}, 114024
  (2017)}.

\bibitem{Chen:2017gnu}
B.~Chen and X.~Liu,
New $\Omega_c^0$ baryons discovered by LHCb as the members of $1P$ and $2S$
  states,
\href{https://doi.org/10.1103/PhysRevD.96.094015}{Phys. Rev. D {\bf 96}, 094015
  (2017)}.

\bibitem{Aliev:2017led}
T.~M. Aliev, S.~Bilmis, and M.~Savci,
Are the new excited $\Omega_c$ baryons negative parity states?
\href{https://doi.org/10.1142/S0217732319503449}{Mod. Phys. Lett. A {\bf 35},
  1950344 (2020)}.

\bibitem{Agaev:2017lip}
S.~S. Agaev, K.~Azizi, and H.~Sundu,
Interpretation of the new $\Omega_c^{0}$ states via their mass and width,
\href{https://doi.org/10.1140/epjc/s10052-017-4953-z}{Eur. Phys. J. C {\bf 77},
  395 (2017)}.

\bibitem{Luo:2021dvj}
S.~Q. Luo, B.~Chen, X.~Liu, and T.~Matsuki,
Predicting a new resonance as charmed-strange baryonic analog of
  $D^*_{s0}$(2317),
\href{https://doi.org/10.1103/PhysRevD.103.074027}{Phys. Rev. D {\bf 103},
  074027 (2021)}.

\bibitem{Zhang:2025gar}
Y.~Zhang, Q.~F. Song, Q.~F. L{\"u}, H.~Nagahiro, and A.~Hosaka,
Understanding the low-lying $\Omega_c$ structures from a coupled-channel
  perspective,
\href{https://doi.org/10.1103/grxl-nwwy}{Phys. Rev. D {\bf 112}, 034035
  (2025)}.

\bibitem{Heikkila:1983wd}
K.~Heikkila, S.~Ono, and N.~A. Tornqvist,
HEAVY c anti-c AND b anti-b QUARKONIUM STATES AND UNITARITY EFFECTS,
\href{https://doi.org/10.1103/PhysRevD.29.2136}{Phys. Rev. D {\bf 29}, 110
  (1984)},
[Erratum: Phys.Rev.D 29, 2136 (1984)].

\bibitem{Ono:1983rd}
S.~Ono and N.~A. Tornqvist,
Continuum Mixing and Coupled Channel Effects in $c \bar{c}$ and $b \bar{b}$
  Quarkonium,
\href{https://doi.org/10.1007/BF01558041}{Z. Phys. C {\bf 23}, 59 (1984)}.

\bibitem{Ono:1985jt}
S.~Ono, A.~I. Sanda, N.~A. Tornqvist, and J.~Lee-Franzini,
Where Are the $B \bar{B}$ Mixing Effects Observable in the $\Upsilon$ Region?
\href{https://doi.org/10.1103/PhysRevLett.55.2938}{Phys. Rev. Lett. {\bf 55},
  2938 (1985)}.

\bibitem{Ono:1985eu}
S.~Ono, A.~I. Sanda, and N.~A. Tornqvist,
$B$ Meson Production Between the $\Upsilon(4S)$ and $\Upsilon(6S)$ and the
  Possibility of Detecting $B \bar{B}$ Mixing,
\href{https://doi.org/10.1103/PhysRevD.34.186}{Phys. Rev. D {\bf 34}, 186
  (1986)}.

\bibitem{Tornqvist:1984fy}
N.~A. Tornqvist and P.~Zenczykowski,
Ground State Baryon Mass Splittings From Unitarity,
\href{https://doi.org/10.1103/PhysRevD.29.2139}{Phys. Rev. D {\bf 29}, 2139
  (1984)}.

\bibitem{Silvestre-Brac:1991qqx}
B.~Silvestre-Brac and C.~Gignoux,
Unitary effects in spin orbit splitting of P wave baryons,
\href{https://doi.org/10.1103/PhysRevD.43.3699}{Phys. Rev. D {\bf 43},
  3699--3708 (1991)}.

\bibitem{Pennington:2007xr}
M.~R. Pennington and D.~J. Wilson,
Decay channels and charmonium mass-shifts,
\href{https://doi.org/10.1103/PhysRevD.76.077502}{Phys. Rev. D {\bf 76}, 077502
  (2007)}.

\bibitem{Barnes:2007xu}
T.~Barnes and E.~S. Swanson,
Hadron loops: General theorems and application to charmonium,
\href{https://doi.org/10.1103/PhysRevC.77.055206}{Phys. Rev. C {\bf 77}, 055206
  (2008)}.

\bibitem{Zhou:2011sp}
Z.~Y. Zhou and Z.~Xiao,
Hadron loops effect on mass shifts of the charmed and charmed-strange spectra,
\href{https://doi.org/10.1103/PhysRevD.84.034023}{Phys. Rev. D {\bf 84}, 034023
  (2011)}.

\bibitem{Danilkin:2010cc}
I.~V. Danilkin and Y.~A. Simonov,
Dynamical origin and the pole structure of X(3872),
\href{https://doi.org/10.1103/PhysRevLett.105.102002}{Phys. Rev. Lett. {\bf
  105}, 102002 (2010)}.

\bibitem{Liu:2016wxq}
Z.~W. Liu, J.~M.~M. Hall, D.~B. Leinweber, A.~W. Thomas, and J.~J. Wu,
Structure of the $\mathbf{\Lambda(1405)}$ from Hamiltonian effective field
  theory,
\href{https://doi.org/10.1103/PhysRevD.95.014506}{Phys. Rev. D {\bf 95}, 014506
  (2017)}.

\bibitem{Zhang:2009bv}
O.~Zhang, C.~Meng, and H.~Q. Zheng,
Ambiversion of $X(3872)$,
\href{https://doi.org/10.1016/j.physletb.2009.09.033}{Phys. Lett. B {\bf 680},
  453--458 (2009)}.

\bibitem{Ortega:2009hj}
P.~G. Ortega, J.~Segovia, D.~R. Entem, and F.~Fernandez,
Coupled channel approach to the structure of the $X(3872)$,
\href{https://doi.org/10.1103/PhysRevD.81.054023}{Phys. Rev. D {\bf 81}, 054023
  (2010)}.

\bibitem{Li:2009ad}
B.~Q. Li, C.~Meng, and K.~T. Chao,
Coupled-Channel and Screening Effects in Charmonium Spectrum,
\href{https://doi.org/10.1103/PhysRevD.80.014012}{Phys. Rev. D {\bf 80}, 014012
  (2009)}.

\bibitem{Kalashnikova:2005ui}
Y.~S. Kalashnikova,
Coupled-channel model for charmonium levels and an option for $X(3872)$,
\href{https://doi.org/10.1103/PhysRevD.72.034010}{Phys. Rev. D {\bf 72}, 034010
  (2005)}.

\bibitem{Danilkin:2009hr}
I.~V. Danilkin and Y.~A. Simonov,
Channel coupling in heavy quarkonia: Energy levels, mixing, widths and new
  states,
\href{https://doi.org/10.1103/PhysRevD.81.074027}{Phys. Rev. D {\bf 81}, 074027
  (2010)}.

\bibitem{Lu:2017hma}
Y.~Lu, M.~N. Anwar, and B.~S. Zou,
How Large is the Contribution of Excited Mesons in Coupled-Channel Effects?
\href{https://doi.org/10.1103/PhysRevD.95.034018}{Phys. Rev. D {\bf 95}, 034018
  (2017)}.

\bibitem{Anwar:2018yqm}
M.~N. Anwar, Y.~Lu, and B.~S. Zou,
$\chi_{b}(3P)$ multiplet revisited: Hyperfine mass splitting and radiative
  transitions,
\href{https://doi.org/10.1103/PhysRevD.99.094005}{Phys. Rev. D {\bf 99}, 094005
  (2019)}.

\bibitem{Ortega:2016pgg}
P.~G. Ortega, J.~Segovia, D.~R. Entem, and F.~Fern{\'a}ndez,
Threshold effects in P-wave bottom-strange mesons,
\href{https://doi.org/10.1103/PhysRevD.95.034010}{Phys. Rev. D {\bf 95}, 034010
  (2017)}.

\bibitem{Ortega:2021fem}
P.~G. Ortega, J.~Segovia, D.~R. Entem, and F.~Fernandez,
The $D_{s0}(2590)^+$ as the dressed $c\bar{s}(2^1S_0)$ meson in a
  coupled-channels calculation,
\href{https://doi.org/10.1016/j.physletb.2022.136998}{Phys. Lett. B {\bf 827},
  136998 (2022)}.

\bibitem{Ortega:2021yis}
P.~G. Ortega, D.~R. Entem, and F.~Fern{\'a}ndez,
Does the $J^{PC}=1^{+-}$ counterpart of the $X(3872)$ exist?
\href{https://doi.org/10.1016/j.physletb.2022.137083}{Phys. Lett. B {\bf 829},
  137083 (2022)}.

\bibitem{Lu:2016mbb}
Y.~Lu, M.~N. Anwar, and B.~S. Zou,
Coupled-Channel Effects for the Bottomonium with Realistic Wave Functions,
\href{https://doi.org/10.1103/PhysRevD.94.034021}{Phys. Rev. D {\bf 94}, 034021
  (2016)}.

\bibitem{Luo:2023sra}
S.~Q. Luo and X.~Liu,
Newly observed $\Omega_c(3327)$: A good candidate for a D-wave charmed baryon,
\href{https://doi.org/10.1103/PhysRevD.107.074041}{Phys. Rev. D {\bf 107},
  074041 (2023)}.

\bibitem{Luo:2023sne}
S.~Q. Luo and X.~Liu,
Investigating the spectroscopy behavior of undetected $1F$-wave charmed
  baryons,
\href{https://doi.org/10.1103/PhysRevD.108.034002}{Phys. Rev. D {\bf 108},
  034002 (2023)}.

\bibitem{Peng:2024pyl}
Y.~X. Peng, S.~Q. Luo, and X.~Liu,
Refining radiative decay studies in singly heavy baryons,
\href{https://doi.org/10.1103/PhysRevD.110.074034}{Phys. Rev. D {\bf 110},
  074034 (2024)}.

\bibitem{Zhang:2025vtc}
Z.~L. Zhang and S.~Q. Luo,
Spectroscopic properties of $1F$-wave singly bottom baryons,
\href{https://doi.org/10.1103/y6gk-2dcw}{Phys. Rev. D {\bf 112}, 074020
  (2025)}.

\bibitem{Hiyama:2003cu}
E.~Hiyama, Y.~Kino, and M.~Kamimura,
Gaussian expansion method for few-body systems,
\href{https://doi.org/10.1016/S0146-6410(03)90015-9}{Prog. Part. Nucl. Phys.
  {\bf 51}, 223--307 (2003)}.

\bibitem{Micu:1968mk}
L.~Micu,
Decay rates of meson resonances in a quark model,
\href{https://doi.org/10.1016/0550-3213(69)90039-X}{Nucl. Phys. B {\bf 10},
  521--526 (1969)}.

\bibitem{LeYaouanc:1972vsx}
A.~Le~Yaouanc, L.~Oliver, O.~Pene, and J.~C. Raynal,
Naive quark pair creation model of strong interaction vertices,
\href{https://doi.org/10.1103/PhysRevD.8.2223}{Phys. Rev. D {\bf 8}, 2223--2234
  (1973)}.

\bibitem{Chen:2016iyi}
B.~Chen, K.~W. Wei, X.~Liu, and T.~Matsuki,
Low-lying charmed and charmed-strange baryon states,
\href{https://doi.org/10.1140/epjc/s10052-017-4708-x}{Eur. Phys. J. C {\bf 77},
  154 (2017)}.

\bibitem{CLEO:1998wvk}
C.~P. Jessop {\it et~al}. (CLEO Collaboration),
Observation of two narrow states decaying into $\Xi^+_c$ gamma and $\Xi^0_c$
  $\gamma$,
\href{https://doi.org/10.1103/PhysRevLett.82.492}{Phys. Rev. Lett. {\bf 82},
  492--496 (1999)}.

\bibitem{BaBar:2006pve}
B.~Aubert {\it et~al}. (BaBar Collaboration),
Observation of an excited charm baryon $\Omega_c^*$ decaying to $\Omega_c^0$
  $\gamma$,
\href{https://doi.org/10.1103/PhysRevLett.97.232001}{Phys. Rev. Lett. {\bf 97},
  232001 (2006)}.

\bibitem{Solovieva:2008fw}
E.~Solovieva {\it et~al}.,
Study of $\Omega_c^{0}$ and $\Omega_c^{*0}$ Baryons at Belle,
\href{https://doi.org/10.1016/j.physletb.2008.12.062}{Phys. Lett. B {\bf 672},
  1--5 (2009)}.

\bibitem{Belle:2020ozq}
J.~Yelton {\it et~al}. (Belle Collaboration),
Study of electromagnetic decays of orbitally excited $\Xi_c$ baryons,
\href{https://doi.org/10.1103/PhysRevD.102.071103}{Phys. Rev. D {\bf 102},
  071103 (2020)}.

\bibitem{BaBar:2003oey}
B.~Aubert {\it et~al}. (BaBar Collaboration),
Observation of a narrow meson decaying to $D_s^+ \pi^0$ at a mass of 2.32
  GeV/c$^2$,
\href{https://doi.org/10.1103/PhysRevLett.90.242001}{Phys. Rev. Lett. {\bf 90},
  242001 (2003)}.

\bibitem{Belle:2025dtr}
M.~Abumusabh {\it et~al}. (Belle, Belle-II Collaboration),
Observation of the Radiative Decay $D_{s0}^*(2317)^+\to D_s^{*+}\gamma$,
\href{https://doi.org/10.1103/vcld-225s}{Phys. Rev. Lett. {\bf 136}, 241901
  (2026)}.

\end{thebibliography}

\end{document}